# Impurity induced Local Magnetism and Density of States in the superconducting state of YBa$_2$Cu$_3$O$_7$

S. Ouazi,[1,*] J. Bobroff,[1] H. Alloul,[1] M. Le Tacon,[1] N. Blanchard,[1] G. Collin,[2]
M.H. Julien,[3] M. Horvatic,[4] C. Berthier[3,4]

[1] Laboratoire de Physique des Solides, UMR8502, Université Paris XI, 91405 Orsay, France
[2] LLB, CE-Saclay, CEA-CNRS, 91191 Gif Sur Yvette, France
[3] Laboratoire de Spectrométrie Physique, Université J. Fourier Grenoble I, BP 87, 38402 Saint-Martin d'Héres, France
[4] Grenoble High Magnetic Field Laboratory, CNRS, BP 166, 38042 Grenoble, France

$^{17}$O NMR is used to probe the local influence of nonmagnetic Zn and magnetic Ni impurities in the superconducting state of optimally doped high T$_C$ YBa$_2$Cu$_3$O$_7$. Zn and Ni induce a staggered paramagnetic polarization, similar to that evidenced above T$_C$, with a typical extension $\xi$=3 cell units for Zn and $\xi \geq 3$ for Ni. In addition, Zn is observed to induce a local density of states near the Fermi Energy in its neighbourhood, which also decays over about 3 cell units. Its magnitude decreases sharply with increasing temperature. This allows direct comparison with the STM observations done in BiSCO.

   Substitution impurities have proved to be a powerfool way to investigate the superconducting and normal state properties of the high T$_C$ cuprates. Initially, most of the experiments in the superconducting state have focused on the average effect of impurities on the macroscopic properties. For example, both the decrease of T$_C$ and the increase of penetration depth induced by non magnetic substitutions at the Cu site of CuO$_2$ planes show that the superconducting gap is anisotropic [1]. The impurity being an atomic size defect, it is essential to monitor its effects on the electronic properties at an atomic scale. In this scope, Scanning Tunneling Microscopy (STM) allows to image the local Density of States (LDOS) induced by an impurity in its immediate vicinity. In conventionnal superconductors, a magnetic impurity has been shown to affect the order parameter over a few lattice cells, as anticipated by BCS theory [2]. In cuprates, similar experiments were performed on Ni and Zn impurities in the superconducting state of BiSCCO. Nonmagnetic Zn affects the superconducting gap and induces LDOS peaks near the Fermi Energy E$_F$, which are seen over a few lattice cells in an anisotropic pattern reminicent of the anisotropy of the d-wave order parameter [3]. In contrast, magnetic Ni does not affect the superconducting gap and induces LDOS peaks far from E$_F$ which were attributed to its own magnetic moment [4]. These effects were anticipated by Cu NMR experiments which showed that the DOS far from defects is increased by Zn, not by Ni [5]. This can be understood in an anisotropic superconducting BCS-model where Zn is assumed to be a much stronger scattering center than Ni, which carries a moment [6]. However, most of these models do not take into account the magnetic correlations which affect the vicinity of the impurity as well. In fact, above T$_C$, NMR studies allowed to show that Zn and Ni induce a staggered paramagnetic polarization in their vicinity, which could be imaged site per site [7,8,9]. This effect is a direct static signature of the magnetic correlations within CuO$_2$ planes. No analogous experimental determination was performed up to now below T$_C$. The only attempt using NMR of the impurity itself suggests that the staggered polarization seen above T$_C$ persists below T$_C$ at least on the impurity first near neighbour Cu sites [10]. NQR T$_1$ measurements studies reveal a concomitent local enhancement of magnetic correlations near the impurity [11]. Even in standard BCS superconductors, no such local measurement of the magnetism nearby an impurity could be achieved to our knowledge.

   In this letter, we measure for the first time the local magnetism around Zn and Ni in the superconducting state using in-plane $^{17}$O NMR. Measurements were performed in high magnetic fields in order to minimize the contribution of the vortex field pattern to the NMR spectra. Both Zn and Ni are found to induce a staggered paramagnetic polarization in continuity with the normal state. In addition, LDOS effects near E$_F$ are observed for Zn, not for Ni, and their evolution with increasing $T$ is measured for the first time.

   We studied optimally doped YBa$_2$Cu$_3$O$_7$ with various in-plane concentrations $x_p$ of Zn and Ni at Cu site. The T$_C$, $\chi$ and $^{17}$O normal state NMR spectra of the same powder samples were already reported [7,9]. The details on the synthesis, impurity substitution control, $^{17}$O enrichment, oxygenation, alignment procedure, are described in these previous reports. Here the $^{17}$O NMR spectra were recorded at external field $H$ in the range 3 T- 14 T and parallel to the c axis. Because of the presence of frozen vortices lattice, broad spectra as shown in Fig.1 were obtained by sweeping the frequency at fixed field $H$ and gluing the Fourier Transform of the echo [12]. We could isolate the in-plane oxygen signal from that of the other oxygen sites by using both time relaxation contrast and measurements on central line together with O(2,3) quadrupolar satellites. Here, in-plane $^{17}$O is better suited than $^{63}$Cu or $^{89}$Y NMR. Indeed, contrary to $^{89}$Y, its hyperfine coupling with in-plane Cu is high enough so that vortex physics is not dominating the lineshape.

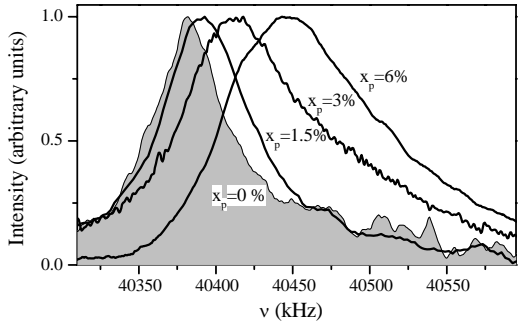

*fig.1 : In plane $^{17}O$ NMR spectra normalized to the maximum intensity for pure (gray) and various Zn in-plane concentrations in $YBaCuO_7$ at H=7 T and T=15K*

But it is small enough to avoid any wipeout of sites near the impurity, as observed for $^{63}Cu$ where large quadrupolar couplings also prohibit sensitive lineshape analysis below $T_C$ [8,11].

Fig.1 shows typical $^{17}O$ NMR spectra below $T_C$. In pure $YBa_2Cu_3O_7$, the linewidth is found to be almost independent of the external field, and similar in size to previous studies done with the same fixed field procedure [13]. The presence of the frozen vortex lattice leads to a field distribution between vortices, hence a frequency distribution in NMR. Such a distribution is predicted to be almost field independant, decreasing by only 10% between 3 T and 14 T [14], consistent with our findings. Upon addition of Zn impurities, a new large paramagnetic contribution overwhelms the vortex field-independent broadening in the largest applied field of 14Tesla. The vortex contribution measured by independent muon spin resonance (µSR) studies, is reported on fig.2. Indeed, the muon spins are not sufficiently coupled to the Copper electronic spins to be sensitive to the Zn paramagnetic contribution. The vortex field distribution is found to decrease with increasing impurity content $x_p$, as the penetration depth λ increases [1,14]. We find that the total NMR broadening is almost linear in field above 5 Tesla and extrapolates at *H=0* to a value which agrees perfectly with that found by muSR, after scaling by the ratio of the gyromagnetic factors of the muon and the $^{17}O$ nucleus [15,16]. This demonstrates that the Zn-induced broadening is proportional to the field, and that the vortices only contribute to a roughly additive broadening in the lowest fields used here. The estimates of the vortex and Zn contributions are given in Fig. 2 for *H=7* T. In contrast with other studies [17], for the fields and concentrations considered, the number of impurities is at least an order of magnitude larger than the number of vortex cores.

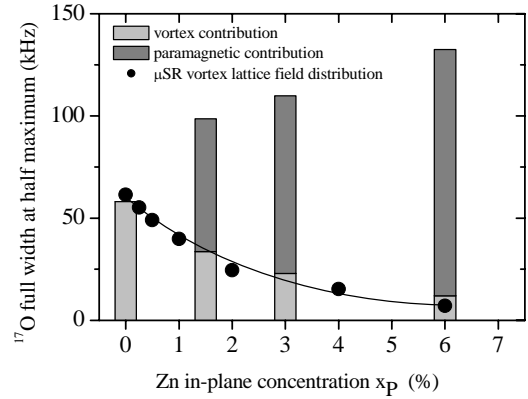

*fig.2 : The vortex (light gray) and the paramagnetic (dark gray) contributions to the full in plane $^{17}O$ NMR width are plotted versus Zn concentration at H=7 T and T=10K. The former contribution agrees quantitatively with the vortex field distribution estimated from independent µSR experiments made at smaller fields on similar compounds [15] (full circles). Solid line is a guide to the eye.*

So only a negligible amount of impurities might pin vortices, and that the observed broadening is dominated by the effect of impurities which are not in vortex cores.

At high fields such as *H=14* T, the Zn contribution is much larger than the vortex one and can then be safely deduced, leading to the results presented in fig.3 and 4 [18]. At low *T*, Zn impurities shift the spectrum towards high frequencies, and broaden the line asymetrically, as observed in fig.1. We checked that the same effects are observed for Li instead of Zn, showing that they are generic of nonmagnetic impurities. The shift of the spectrum is mostly due to an *average* DOS effect and will be addressed later. The broadening shows that Zn induces an additional local paramagnetic distribution. The low and high frequency half widths of this broadening are noted $\Delta\nu_L$ and $\Delta\nu_H$. The former is plotted on fig.3, while the asymmetry of the broadening $\Delta\nu_H$-$\Delta\nu_L$ is plotted on fig.4. $\Delta\nu_L$ increases almost linearly with Zn content and is almost temperature independent. Its value being very close to the one measured at *T*=100K above $T_C$, this broadening must originate from a staggered paramagnetic polarization induced by Zn as in the normal state [9]. Such a staggered cloud persists in the superconducting state, both its extension and its amplitude being almost *T*-independent. The Ni impurity is also observed to induce alternated moments, as above $T_C$ [7]. But the corresponding broadening follows a Curie-Weiss $C/(T+\Theta)$ dependence with $\Theta=39$ K in contrast with Zn. Such difference probably originates from the fact that Zn is spinless, whereas Ni carries a magnetic moment with its own *T*-dependent local susceptibility. Anyhow, the induced cloud displays similar characteristics for both Zn and Ni, as will be demonstrated later.

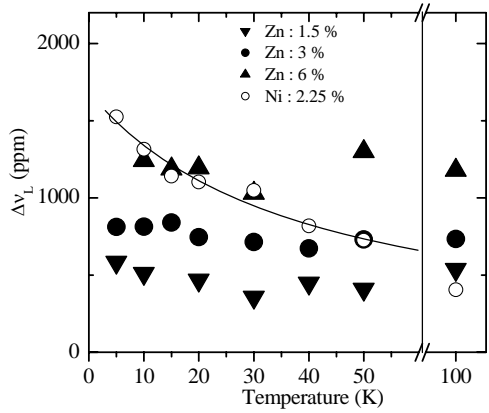 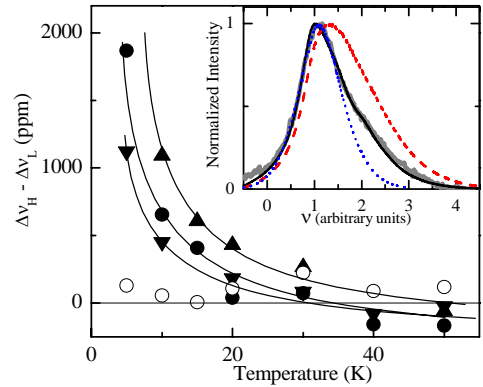

fig.3 : Paramagnetic low frequency half width $\Delta\nu_L$ of the in-plane $^{17}O$ NMR line for Zn and Ni substituted $YBa_2Cu_3O_7$. The Full line is a Curie-Weiss fit of the broadening due to Ni impurity.

fig.4 : Main panel : Paramagnetic asymmetry $\Delta\nu_H - \Delta\nu_L$ of the in-plane $^{17}O$ NMR line for Zn:1.5%, Zn:3%, Zn:6% (full down triangle, circle, up triangle) and Ni:2.25% (empty circle) substituted $YBa_2Cu_3O_7$. Full lines are guide to the eye. The asymmetry is seen to increase for Zn below T=30K but not for Ni. Inset: $^{17}O$ NMR line for Zn:3%, T=10K, H=7 T (gray line) and simulations for a purely alternated model (blue dotted), an alternated+non alternated model with $\xi=3$ and $\lambda=3$ (black) and with $\lambda=3.5$ (red dashed).

Sizable asymmetry measured by $\Delta\nu_H - \Delta\nu_L$ on fig.4 is only observed for Zn, not for Ni. It cannot originate from a simple change in the spatial shape of the staggered cloud as $\Delta\nu_H$ and $\Delta\nu_L$ display different *T*-dependences. We propose an interpretation in terms of a local density of states (LDOS) effect, as observed by STM [3]. In a metal, the NMR « Knight » shift is directly proportionnal to the DOS at $E_F$. Here as well, any impurity-induced DOS close to $E_F$, as the one observed by STM for Zn, should shift the corresponding NMR frequency on the high frequency side. Note that a uniform DOS effect on all Cu sites due to Zn would only homogeneously shift the NMR line but would not affect its width or shape. This is what has been seen and implicitly assumed in previous Cu NMR studies [5]. Only a *space-varying* LDOS which decreases from the impurity site can explain both the increase of $\Delta\nu_H - \Delta\nu_L$ and the shift. In the Ni case, STM studies measured LDOS peaks but far from $E_F$ [4], which should thus have no effect in NMR. This is indeed what we observe for Ni, as the $^{17}O$ NMR line is symetric at all temperatures (fig.4). Furthermore, it is not shifted towards higher frequencies as compared to the pure sample, contrary to the Zn case.

In summary, the low frequency part of the spectrum enables us to probe the staggered moments induced by Zn or Ni, which persist below $T_C$. The asymmetry probes additionnal LDOS effects near $E_F$, which develop only at low *T* for Zn. A quantitative estimate of the spatial extension of both effects can be made following the procedure developed in [9] with the same notations for $\xi$. Using a given model for the polarization induced by each impurity, one can reconstruct the expected NMR lineshape with the known values and form factors for $^{17}O$ hyperfine couplings. Above $T_C$, a staggered polarization with a Bessel enveloppe $K_0(r/\xi)$ fits well all Zn NMR data with an extension $\xi_{imp} \sim 3\pm1$ cell units at $T=100K$ [9]. Below $T_C$, as the low frequency part of the spectrum is unchanged, the same enveloppe fits the data. We account for the LDOS effect by an additional positive decaying component of the spin polarization around each Zn. All simulations are then convoluted with the small asymetric vortex field distribution estimated before. We succeeded in fitting the Zn induced broadening for all $x_p$ at low *T* with an LDOS which decays exponentially as $\exp(-r/\lambda)$, with $\lambda \sim 3$ unit cells. The examples of simulations given in the inset of fig.4 show that a purely alternated model fails to reproduce the experimental asymetry while the abovementionned combined model fits well the data. The sensitivity to the extension $\lambda$ is also demonstrated. Any more refined analysis of the actual shape of the decay, of its possible in-plane anisotropy, or of the *T*-dependence of $\lambda$ is beyond the reliability of our analysis. Our simulations also lead to a high frequency shift of the line, as observed on fig.1. However the experimental shift comes from both the change of the center of gravity of the vortex field distribution [14] and from the paramagnetic shift due to LDOS effects [5]. The latter is very sensitive to the overlap effects between LDOS patterns from different impurities. The average distance between Zn is not high enough here to allow any quantitative comparison.

Overall, our findings agree well with the STM estimates of the LDOS decay on the first atomic layer of BiSCCO [3]. We conclude that STM probes a physics qualitatively similar to our measurements in the bulk of optimally doped YBaCuO. Our data allows us to follow the T variation of the LDOS, never measured before. Indeed the asymmetry of the spectrum decreases sharply with increasing *T*, and becomes undetectable within experimental accuracy above T=30K. We correspondingly expect a broadening and/or a decrease of the LDOS peak with increasing *T*, which might be checked by *T*-dependent STM measurements, not available up to now.

In the case of Ni, the staggered paramagnetic polarization alone allows to fit the spectral width with no additionnal LDOS, as expected. The spectra can be accounted for with $\xi \geq 3$. For smaller $\xi$, simulations yield a large low frequency satellite not observed experimentally. One could naively consider the moments induced by Zn and Ni below $T_C$ as resulting from a mere continuation of the normal state physics in a "normal" region close to the impurity where the superconducting gap is destroyed. However, STM experiments indicate that Ni does not affect the superconducting gap in contrast with Zn. So this paramagnetic staggered cloud coexists locally with superconductivity at least in the Ni case. Previous $^7$Li NMR measurements in the case of the non magnetic Li impurity revealed below $T_C$ an increase of the Curie-Weiss susceptibility on n.n. Cu, interpreted as resulting from a decrease of the normal state Kondo screening [10]. The existence of the staggered moments is confirmed here, but their temperature dependence is much less prononced than anticipated from the $^7$Li NMR data. The Li measurements could be partly affected by the LDOS peak at optimal doping. The estimated Kondo temperature $T_K$=41 K below $T_C$ would then be underestimated, the actual $T_K$ being closer to the normal state value of 135 K.

The LDOS effects observed in STM and in the present study are predicted for strong scattering centers such as Zn or Li in a d-wave superconductor [6]. However, the BCS type models used in ref [6] do not take into account the staggered magnetic effects. Models based on strong magnetic correlations allow to account for these staggered moments above $T_C$ (see refs. in [9]). Using a t-J model in a d-wave superconductor, Wang and Lee demonstrated that correlations lead simultaneously to staggered moments and LDOS peaks [19]. Another set of models attributes the LDOS peaks to a Kondo resonance [20]. In this scenario, the difference between the *T*-dependence of Ni and Zn on fig.3 could originate from a much lower $T_K$ for Ni than for Zn. But the origin of this difference is still unclear. Models addressing both Ni and Zn cases are therefore needed to fulfill our understanding.

In conclusion, $^{17}$O NMR allowed us for the first time to measure the local magnetic and DOS effects due to a nonmagnetic Zn and a magnetic Ni impurity in a superconductor. This is the first microscopic characterization of an impurity-induced magnetism in the superconducting state of any cuprate. The ability to measure LDOS effects as well opens a new way to simultaneously probe magnetic and charge effects near local defects at the atomic scale in the bulk of superconductors. It can be extended to other dopings or materials in a more controled way than the corresponding STM measurements.

*We acknowledge K. Van Der Beek and P. Mendels for fruitful dicsussions, H. Brune for his comprehension.*